\documentclass[aps,prd,reprint,superscriptaddress,nofootinbib,floatfix,longbibliography]{revtex4-2}

\usepackage{amsmath}
\usepackage{bm}
\usepackage{amsfonts}
\usepackage{graphicx}
\usepackage{hyperref}
\usepackage{soul}
\usepackage{color}
\usepackage{physics}

\enlargethispage{\baselineskip}

\hypersetup{
     colorlinks   = true,
     citecolor    = blue
}
\everymath{\displaystyle}

\begin{document}

\title{The anomalous shift of the weak boson mass and the quintessence electroweak axion}

\newcommand{\TDLI}{\affiliation{Tsung-Dao Lee Institute (TDLI) \& School of Physics and Astronomy, Shanghai Jiao Tong University, \\ Shengrong Road 520, 201210 Shanghai, P.\ R.\ China}}

\author{Weikang Lin}
\email{weikanglin@sjtu.edu.cn}
\TDLI

\author{Tsutomu T. Yanagida}
\email{tsutomu.tyanagida@sjtu.edu.cn}
\TDLI
\affiliation{Kavli IPMU (WPI), The University of Tokyo, Kashiwa, Chiba 277-8583, Japan}

\author{Norimi Yokozaki}
\email{n.yokozaki@gmail.com}
\affiliation{
Zhejiang Institute of Modern Physics and Department of Physics, Zhejiang University, Hangzhou, Zhejiang 310027, China}

\date{\today}

\begin{abstract}
One of the simplest ways to account for the observed W-boson mass shift is to introduce the $SU(2)_L$  triplet Higgs boson with zero hypercharge, whose vacuum expectation value is about 3 GeV. If the triplet is heavy enough at $\mathcal{O}(1)$ TeV, it essentially contributes only to $T$ parameter without any conflict to the observation. The presence of a complex triplet Higgs boson raises the $SU(2)_L$ gauge coupling constant to $\alpha_2(M_{\rm PL} )\simeq 1/44$ at the Planck scale. Thanks to this larger gauge coupling constant, we show that the electroweak axion vacuum energy explains the observed cosmological constant provided that the axion field is located near the hill top of the potential at present.
\end{abstract}

\maketitle

\section{The $W$ boson mass shift and the weak SU(2) gauge coupling constant}
The CDF collaboration recently reported an updated precise measurement of the W-boson mass~\cite{CDF:2022hxs}, which is significantly larger than the Standard Model (SM) prediction and 
the world average without this new CDF result~\cite{Workman:2022ynf}.  
With this new CDF result, the experimental average of the W-boson mass becomes~\cite{deBlas:2022hdk}
\begin{eqnarray}
(M_W)_{\rm exp} =(80.4133 \pm 0.0080)\, {\rm GeV}, \label{eq:ave_exp}
\end{eqnarray}
which deviates from the SM prediction~\cite{deBlas:2022hdk} by 6.5 $\sigma$:
\begin{eqnarray}
(M_W)_{\rm SM} = (80.3499 \pm 0.0056)\, {\rm GeV}. \label{eq:sm_pre}
\end{eqnarray}

The discrepancy, $\delta M_W \equiv (M_W)_{\rm exp} - (M_W)_{\rm SM} \sim 60$\,MeV 
is easily explained by the small vacuum expectation value (VEV) of a $SU(2)_L$ triplet Higgs boson with zero hypercharge~\cite{Strumia:2022qkt, DiLuzio:2022xns, Athron:2022isz, FileviezPerez:2022lxp, Evans:2022dgq}, which only contributes to the $W^\pm$ mass at the tree-level~\cite{Ross:1975fq, Gunion:1989ci, Blank:1997qa, Forshaw:2003kh, Chen:2006pb, Chankowski:2006hs, Chivukula:2007koj}. The mass shift of the W-boson may suggest the existence of the triplet Higgs boson at the TeV scale. 

Let us discuss the W-boson mass shift, $\delta M_W \sim 60\,{\rm MeV}$, with the complex $SU(2)_L$ triplet Higgs boson, $\Sigma$, added to the SM Higgs potential. We consider the following potential:
\begin{eqnarray}
 V &\ni&  - m_H^2 |H|^2 + \lambda_H |H|^4 \nonumber \\
  &+& 2  m_{\Sigma}^2 {\rm Tr } \Sigma^\dag \Sigma + (A_{H} H^\dag \Sigma H + h.c.) + V_4
  , \label{eq:potential_triplet}
\end{eqnarray}
where we write
\begin{eqnarray}
	\Sigma = \Sigma^a T^a = \frac{1}{2}
	\left(
	\begin{array}{cc}
		\Sigma^3	&  \sqrt{2} X_1^+ \\
		\sqrt{2} X_2^- & -\Sigma^3
	\end{array}
	\right).
\end{eqnarray}
Here, $T^a$ is a generator of $SU(2)_L$, which satisfies ${\rm Tr}(T^a T^b) = 1/2 \delta^{ab}$; we assume only $H^\dag \Sigma H$ term violates a global $U(1)$ symmetry, $\Sigma \to e^{i \alpha}\Sigma$; we consider the case where $m_{\Sigma}^2 > 0$ and $m_{\Sigma}^2$ is much larger than the squared of the weak scale; $V_4$ contains the quartic terms, $\Tr(\Sigma^\dag \Sigma)^2$, $\Tr(\Sigma^\dag \Sigma^\dag) \Tr(\Sigma \Sigma)$, $\Tr(\Sigma^\dag \Sigma \Sigma^\dag \Sigma)$, $\Tr(\Sigma^\dag \Sigma^\dag \Sigma \Sigma)$.\footnote{Here, we assume $V_4$ respects a $U(1)$ symmetry: $\Sigma \to e^{i \alpha}\Sigma$.} However, these terms are not important in our case. By putting $\left< H \right>=(0,v)^T$, the potential for $\Sigma^3$ near the origin is dominated by the (effective) linear term and the quardratic term, $-(A_H v^2 \Sigma^3/2 +h.c.) + m_{\Sigma}^2 |\Sigma^3|^2$. The quartic terms become important only for $\Sigma^3 \sim m_{\Sigma}$, which is irrelevant in our case. From these facts, we neglect $V_4$ in the following discussions.
%
%
%

By minimizing the potential in Eq.~\eqref{eq:potential_triplet}, the vacuum expectation values for the doublet and triplet Higgs are
\begin{eqnarray}
v^2 =(m_H^2 + A_{H} v_T)/(2\lambda_H), \ \ 
v_T =  \frac{A_{H} v^2}{2 m_{\Sigma}^2},
\end{eqnarray}
where 
\begin{eqnarray}
\left< H \right> = (0, v)^T, \ 
\left< \Sigma \right> = \frac{1}{2}
\left(
\begin{array}{cc}
 v_T	&  0 \\
 0	& -v_T
\end{array}
\right)	
\end{eqnarray}

At the tree-level, the W-boson mass with the triplet Higgs is expressed as (see e.g.,~\cite{Diessner:2019ebm})
\begin{eqnarray}
M_W &\approx&  (M_W)_{\rm SM} \left[ 1 + \frac{1}{2} \frac{\cos^2 \theta_W}{\cos^2 \theta_W-\sin^2 \theta_W} (\alpha T_{\rm tree})  \right], \nonumber \\ 
\end{eqnarray}
where $\theta_W$ is the weak mixing angle, $\alpha$ is the fine structure constant and $T_{\rm tree}$ is the $T$-parameter at the tree-level:
\begin{eqnarray}
T_{\rm tree} = \frac{4 v_T^2}{v^2} \frac{1}{\alpha}.	
\end{eqnarray}
As shown in Refs.~\cite{Strumia:2022qkt, deBlas:2022hdk} (see also \cite{Lu:2022bgw}), $T \sim 0.15, S=0$ can explain $\delta M_W$. It corresponds to $v_T \sim 3$\,GeV in our case.
The loop effects to the oblique parameters, $S$ and $T$, are suppressed as $\sim k^2/16\pi^2 (M_W^2/m_{\Sigma}^2)$ ($k$ is a relevant coupling), which are negligible for $m_{\Sigma} \geq 1\,{\rm TeV}$. For the $U$ parameter, since it comes from a dimension eight operator, the effect is even more suppressed. So far, only $T \simeq T_{\rm tree}$ is non-negligible.

\begin{figure}[htp]
	\centering
	\includegraphics[scale=0.52]{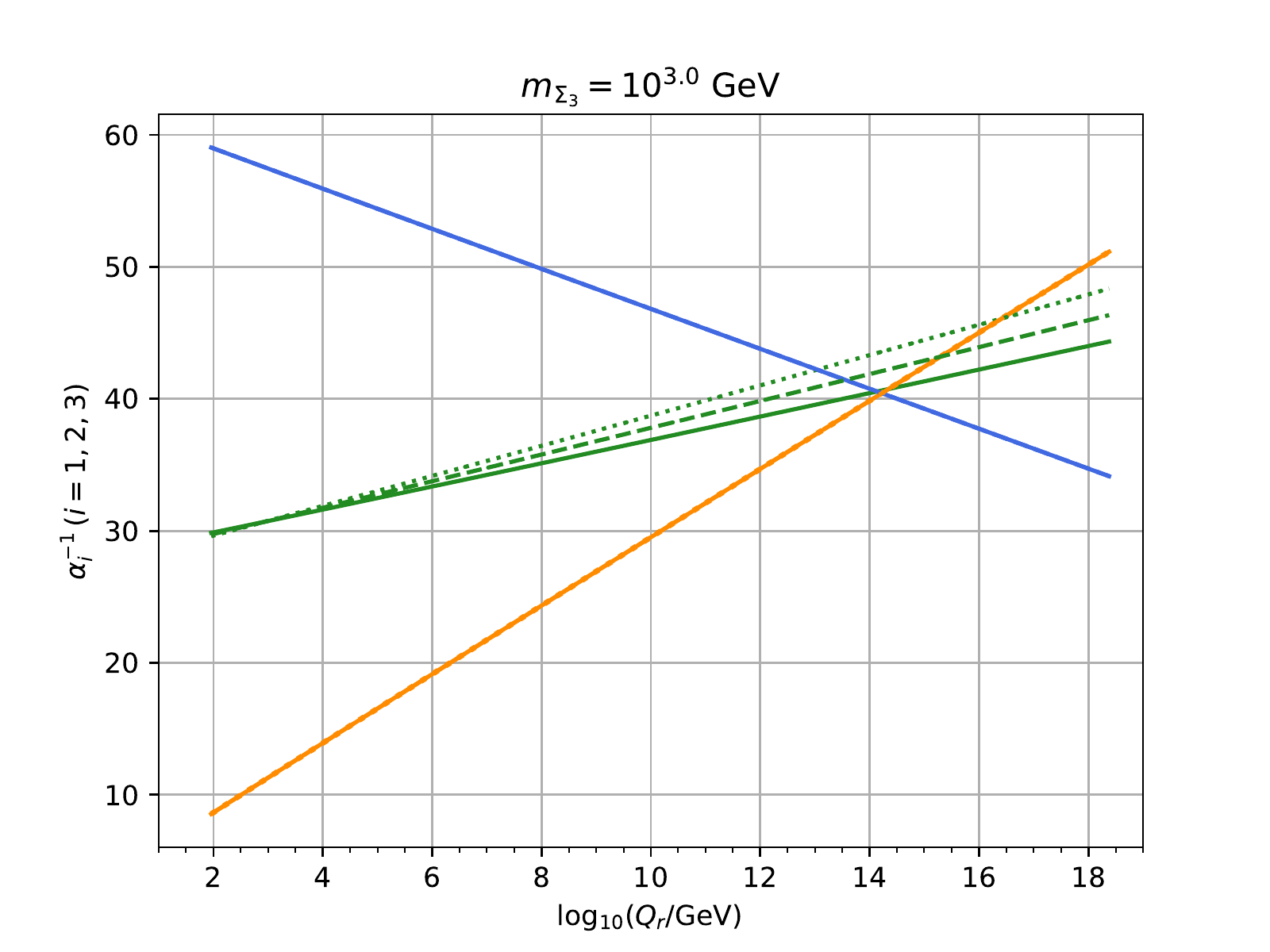}
	\includegraphics[scale=0.52]{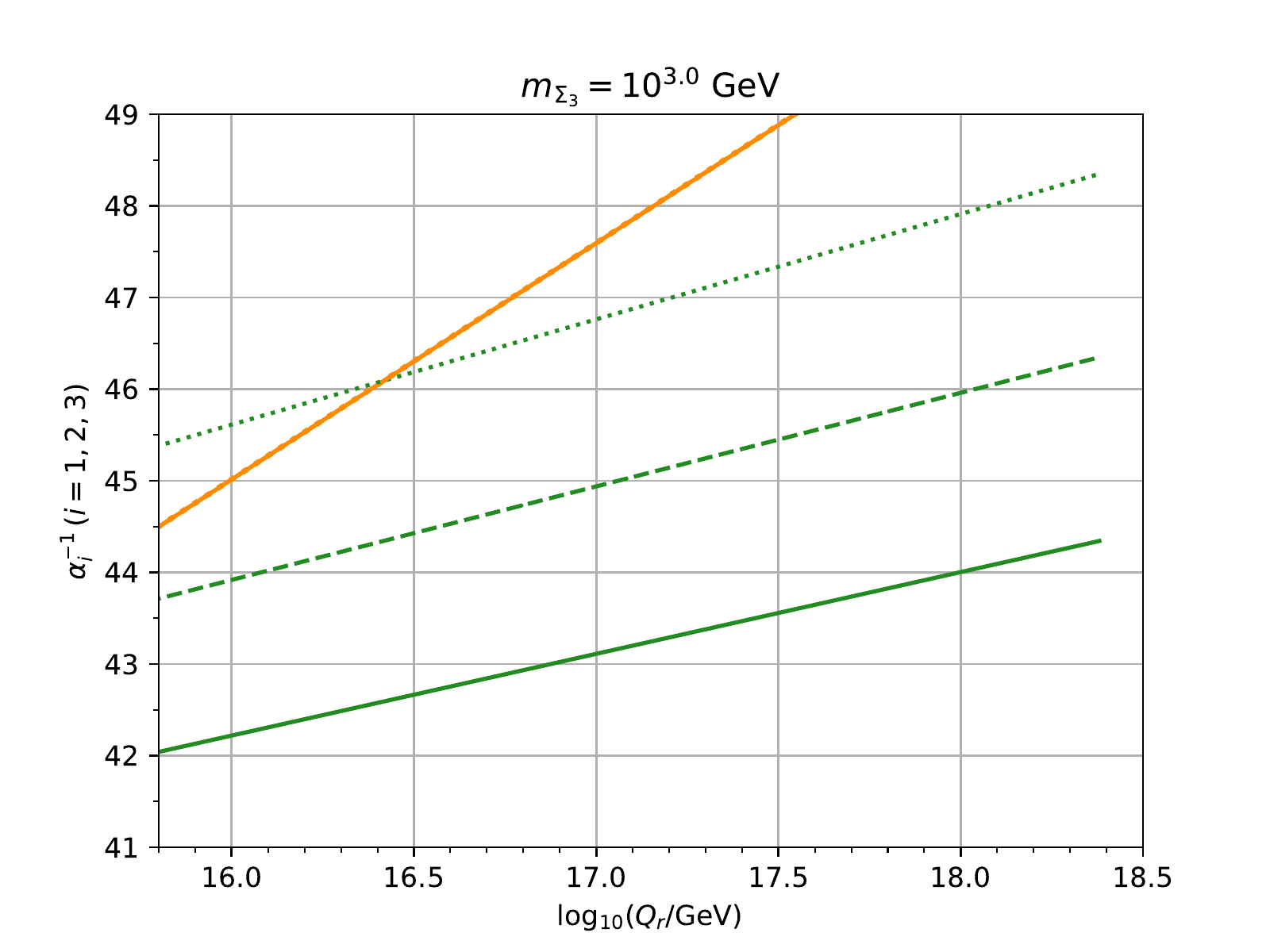}
	\caption{Renormalization group evolution of gauge coupling constants in SM (dotted), SM + one real triplet Higgs (dashed) and SM + one complex triplet Higgs (solid). The blue, green and orange lines show $U(1)_Y$, $SU(2)_L$ and $SU(3)_C$ couplings, respectively. (Here, $\alpha_i^{-1}=4\pi/g_i^2$). The mass of the triplet Higgs is taken as 1\,TeV.
	The bottom plot is the zoomed one of the top plot.} 
	\label{fig:1}
\end{figure}

The mixing between the SM Higgs and the triplet Higgs comes from the scalar trilinear coupling in Eq.\eqref{eq:potential_triplet}. The trilinear coupling $A_H$ is related to $m_\Sigma$ as
\begin{eqnarray}
\frac{A_H}{m_\Sigma} = \frac{2 v_T m_\Sigma}{v^2} \approx 0.3 
\left(
\frac{m_\Sigma}{1500\,{\rm GeV}}	
\right)
\left(
\frac{v_T}{3\,{\rm GeV}}	
\right). \label{eq:ratio_ah}
\end{eqnarray}
The ratio, $A_H/m_{\Sigma}$, increases linearly with $m_{\Sigma}$. This relation is almost unchanged even if quartic couplings involving $\Sigma$ are included (unless $m_\Sigma^2 < 0$). The CP-even Higgs bosons in $H$ and $\Sigma^3$ are mixed due to the third term in Eq.~\eqref{eq:potential_triplet}: 
\begin{eqnarray}
	\left(
	\begin{array}{c}
		h\\
		\sigma_H
	\end{array}
	\right)=
	\left(
	\begin{array}{cc}
	\cos\theta_{\rm mix}	&  \sin\theta_{\rm mix}\\
	-\sin\theta_{\rm mix}	&\cos\theta_{\rm mix}
	\end{array}
	\right)
	\left(
\begin{array}{c}
	\sigma_0 \\
	\sigma_3
\end{array}
\right),
\end{eqnarray}
where $\sigma_0/\sqrt{2} \in H$ and $\sigma_3/\sqrt{2} \in \Sigma^3$; $h$ and $\sigma_H$ are the mass eigenstates: they correspond to the SM-like Higgs and the CP-even heavy Higgs, respectively. The masses of these Higgs bosons are~\cite{Evans:2022dgq}~\footnote{For the other Higgs bosons, see Ref.~\cite{Evans:2022dgq}.} 
\begin{eqnarray}
	m_{h}^2 &\simeq& 4 \lambda_H v^2 - \frac{A_{H}^2 v^2 }{m_{\Sigma}^2} + \mathcal{O}(v^4)
	\nonumber \\ 
	m_{\sigma_H}^2 &\simeq& m_{\Sigma}^2 + \frac{A_{H}^2 v^2 }{m_{\Sigma}^2} + \mathcal{O}(v^4).
\end{eqnarray}
The mixing angle, $\theta_{\rm mix}$, is given by
\begin{eqnarray}
\tan 2 \theta_{\rm mix} &=& \frac{2 A_{H} v}{m_{\Sigma}^2 - 4 \lambda_H v^2} \nonumber \\
&\approx&  
\frac{2 A_{H}}{m_\Sigma} \frac{v}{m_\Sigma}
\approx
0.07
\left(
\frac{v_T}{3\,{\rm GeV}}	
\right).
\end{eqnarray}
The mixing angle is almost fixed by the VEV of the triplet as long as $m_{\Sigma} \gg v$, with $\theta_{\rm mix} \approx 0.035$.   
Therefore, the SM-like CP-even Higgs couplings are modified by a fixed amount ($\theta^2_{\rm mix}/2 \approx 0.06\%$) compared to those of the SM Higgs in the heavy triplet region, which is a kind of general prediction.

On the other hand, when $A_H/m_\Sigma \gtrsim 1$, it induces a deeper global minimum other than the electroweak symmetry breaking minimum even if we include quartic couplings containing $\Sigma$.\footnote{For instance, one can consider the direction $\left<\sigma_0\right>=\left<\sigma_3\right>=x\  (\sigma_0 \in H, \sigma_3 \in \Sigma^3)$ in the scalar potential. Then, the potential can be written as $V_{\rm eff} \simeq m_{\Sigma}^2 x^2 - 2 A_{3 H} x^3 + l_{\rm eff} x^4$, which has the deeper global minimum for $A_{3 H}^2/l_{\rm eff} > (8/9) m_{\Sigma}^2$.} We expect the vacuum stability puts the upper-bound on $m_{\Sigma}$, which is not much larger than 5\,TeV.

\vspace{20pt}
The presence of the complex $SU(2)_L$ triplet with the zero hypercharge at the TeV scale changes the renormalization group evolution of the gauge coupling constants. In particular, this triplet Higgs only affects the renormalization group equation (RGE) for the $SU(2)_L$ gauge coupling at the one-loop level:
\begin{eqnarray}
\frac{d \alpha_2^{-1}}{d \ln Q_r} = -\frac{1}{2\pi} \left(
b_{\rm SM} + \frac{2}{3}
\right), \label{eq:a2beta}
\end{eqnarray}
where $Q_r$ is the renormalization scale and $b_{\rm SM}=-19/6$. 
In Fig.~\ref{fig:1}, the RG evolution of the gauge coupling constants at the two-loop level are shown. We use {\tt PyR@TE 3}~\cite{Sartore:2020gou} to obtain the two-loop RGEs. 
Interestingly, as shown in Ref.~\cite{Evans:2022dgq} and Fig.~\ref{fig:1}, the complex $SU(2)_L$ triplet with the mass of $\sim 1$\,TeV leads to rather precise gauge coupling unification. This suggests us to consider a grand unified theory (GUT) at the scale around $10^{14}$\,GeV. It has been shown that $SU(5)$ GUT can successfully be constructed without inducing rapid proton decays~\cite{Evans:2022dgq}.  

\section{The quintessence electroweak axion dark energy }

Now what if there were no GUT? Surprisingly, in this case, the observed dark energy may be explained by the potential of the electroweak (EW) axion \cite{Fukugita:1994xx}.\footnote{A proposal to identify the quintessence for the dark energy with a pseudo Nambu–Goldstone boson (axion) is first discussed in \cite{Fukugita:1994xx}, \cite{Frieman:1995pm}, \cite{Carroll:1998zi} and \cite{Choi:1999xn}.}
It was pointed out that the $SU(2)_L$ instantons give a correct magnitude of the observed dark energy in the supersymmetric (SUSY) standard model~\cite{Nomura:2000yk}. 

First, let us discuss a short summary of the SUSY EW axion.
We assume that a massless Nambu-Goldstone boson (i.e. EW axion), $A$, couples to the $SU(2)_L$ gauge field as
\begin{equation}\label{eq:AWW-coupling}
    \frac{g^2_2}{32\pi^2} \frac{A}{F_A} W_{\mu \nu}^i\widetilde{W}^{i\mu \nu},
\end{equation}
where $F_A$ is the decay constant of the Nambu-Goldstone boson (which is determined by an  ultraviolet theory).
Integrating out the $SU(2)_L$ instantons, we obtain the following EW axion potential where $\bm W$ and $\bm{\widetilde{W}}$ are the electroweak gauge field and its dual~\cite{Nomura:2000yk}
\begin{equation}
   V=\frac{1}{2}\Lambda_A^4 (1-\cos(A/F_A)), 
\end{equation}
where
\begin{equation}\label{eq:potential-EW-instanton-SUSY}
    \Lambda_A^4 = 2c\times \epsilon^{10} m_{3/2}^3 M_{\rm PL} e^{-\frac{2\pi}{\alpha_2(M_{\rm PL})}}.
\end{equation}
Here, $c$ is an $\mathcal{O}(1)$ constant, $\epsilon\simeq 1/17$ \cite{Buchmuller:1998zf} the flavor suppression factor due to the Froggatt-Nielsen symmetry \cite{Froggatt:1978nt} and $\alpha_2(M_{\rm PL}) \simeq 1/23$ the $SU(2)_L$ gauge coupling constant  at the Planck scale $M_{\rm PL}\simeq 2.4\times 10^{18}$ GeV. We reproduce the observed dark energy at around the hill top of the EW axion potential, that is \cite{Planck:2018vyg, Workman:2022ynf}
\begin{equation}\label{eq:CC-observed}
    \Lambda_{\rm obs}^4 \approx 7 \times 10^{-121} M_{\rm PL}^4,
\end{equation}
for $c\simeq 1$ and the gravitino mass $m_{3/2}\simeq1.8$ TeV. 

However, in the non-SUSY SM, the $SU(2)_L$ gauge coupling constant is smaller as $\alpha_2(M_{\rm PL}) \simeq 1/48$, and hence the potential energy of the EW axion becomes too small. This can be partially remedied by removing the flavor suppression factor $\epsilon^{10}$. For instance, a model with a discrete $Z_{10}$ Froggatt-Nielsen symmetry can generate the axion potential without the flavor suppression factor~\cite{Choi:2019jck}. 
This $Z_{10}$ Froggatt-Nielsen symmetry is anomaly free and may originate from a gauge  symmetry~\cite{Choi:2019jck}. %
Alternatively, we can simply consider cases without the flavor symmetry in the non SUSY standard model. 
Then, the maximal potential energy is given by~\cite{Ibe:2018ffn}
\begin{equation}\label{eq:potential-nonSUSY-Ibe}
    \Lambda_A^4 = c'\times M_{\rm PL}^4 e^{-\frac{2\pi}{\alpha_2(M_{\rm PL})}},
\end{equation}
which still gives the \textbf{ten} orders of magnitude smaller than the observed dark energy in Eq.~\eqref{eq:CC-observed}.\footnote{It is pointed out that we can explain the observed vacuum energy if we adopt a larger cut-off scale than the Planck scale $M_{\rm PL}$ and c' a larger than $\mathcal{O}(1)$ \cite{McLerran:2012mm}.}

Now we return to the case of SM + one complex $SU(2)_L$ triplet Higgs. 
From Eq.~\eqref{eq:a2beta}, it is clear that $\alpha_2^{-1}(M_{\rm PL})$ becomes larger than that in SM. The shift is estimated as $-1/(3\pi) \ln (M_{\rm PL}/1{\rm TeV}) \approx -3.8$, which gives $\exp(2\pi  (3.8)) \approx 2.3 \times 10^{10}$. Therefore, the potential energy of the EW axion explains the observation Eq.~(\ref{eq:CC-observed}) at $A\simeq \pi F_A$.

\section{Discussion}

In all of the above discussions on the potential energy of the EW axion, we have assumed that the vacuum energy of the true potential minimum vanishes. This is at least consistent with the de-Sitter conjecture on the quantum breaking of the classical gravity \cite{Dvali:2014gua} or on the string swampland \cite{Vafa:2005ui}, or our EW axion dark energy hypothesis is consistent with the anthropic principle \cite{Weinberg:2000yb}.\footnote{The dark energy at the present universe may consist of two components unless the axion sits at its potential minimum already. One is truly the cosmological constant and the other the potential energy of the EW axion. The total dark energy is then constrained by the anthropic principle \cite{Weinberg:2000yb}. If both terms are comparable and the cosmological constant is positive, we do not have a severe fine-tuning problem discussed below even if the decay constant is $F_A\simeq 10^{16}$ GeV for a good quality of the EW axion. And the cosmic birefringence can be natuarlly explained by the movement of the EW axion. However, this positive cosmological constant is inconsistent with the de-Sitter conjecture. For the model to be consistent with the de-Sitter conjecture we must consider the cosmological constant to be the vacuum energy of some other light field $B$.} But if we want to understand the vanishing cosmological constant dynamically,
we need some deeper insight of quantum gravity which is beyond the scope of this paper.\footnote{If there is some symmetry acting on the wave function of the universe, like the chiral symmetry acting on massless fermionic fields in quantum field theory, it might be possible to impose the cosmological constant to vanish.}

However, even if the quintessence axion hypothesis is interesting, we have two serious problems. One is the quality problem of the shift symmetry as quantum gravity is believed to explicitly break any global symmetries.
The other is the fine tuning problem of the initial axion field value. Namely, we have to assume that the initial value of the EW axion $A_{\rm i}$ is taken near the hill top of the potential, that is, $A_{\rm i}\simeq \pi F_A$. The initial condition would not be a serious problem if $F_A \simeq M_{\rm PL}$. However, if it is the case, the axion quality problem becomes very serious. We describe such a conflict in a more quantitative way as follows.

As for the breaking of the shift symmetry, we have probably many processes in quantum gravity. Let us discuss wormhole instanton effects here, since it can be estimated in a relatively straightforward way by solving wormhole-axion solutions \cite{Giddings:1987cg}. The detailed calculation is given in \cite{Alonso:2017avz} in which the relevant Euclidean action of the wormhole is given by
\begin{equation}\label{eq:Swh}
    S_{wh} \simeq 0.14\times \sqrt{8\pi} \frac{M_{\rm PL}}{F_A}\,.
\end{equation}
Imposing a sufficiently small suppression, i.e., $\exp(-S_{wh})<10^{-120}$, we obtain a constraint on $F_A < 0.7\times 10^{16}$ GeV.

However, we consider that the above result is speculative due to the absence of the detailed knowledge of a quantum gravity. Maybe we have some non-perturbative contributions $\gamma$ \cite{Giddings:1987cg} to the wormhole action $S_{wh}$ which may make the total wormhole action larger as long as no cancellations in different contributions take place. Thus, we consider the decay constant $F_A \simeq 10^{16}$ GeV is sufficiently small to suppress the explicit breaking of the axion shift symmetry by the wormholes.\footnote{For a recent discussion on the quality problem of axions in a linear $\sigma$ model see \cite{Choi:2022fha}.}

On the other hand, in order to keep the equation of state of the axion to be within the observational bound, i.e., $|w+1|<0.05$ \cite{Planck:2018vyg}, the initial $A_I$ needs to be extremely close to the hill top of the potential. From a WKB approximation to the dynamics of the EW axion (and combining Eq. (10) in \cite{Choi:2021aze}), one can show that 
\begin{equation}\label{eq:Initial-Condition}
    \left|\frac{A_I}{F_A} -\pi \right| \le \frac{M_{\rm PL}}{F_A} e^{-\mathcal{O
    }(M_{\rm PL}/F_A)} \,.
\end{equation}
The above is $\sim10^{-4}$ for $F_A\simeq2\times10^{17}$\,GeV but is $\sim 10^{-100}$ for $F_A\simeq10^{16}$\,GeV. Thus, an $F_A$ that is sufficiently small to keep a good quality of the EW axion will need an extreme fine-tuning of the initial condition if the EW axion is the dark energy. 

To avoid the above severe conflict between the EW axion quality and the naturalness of the initial condition, one can either (1) invoke some mechanism to suppress the wormhole effects for the shift symmetry breaking; (2) study some ways to alleviate or eliminate the fine tuning of the initial condition; or (3) assume the dark energy has some other origin and allow the EW axion to be a fraction of dark matter.

A solution achieving the above (1) may be given in higher dimensional space-time where the extra dimensions are compactified. In this case, the above wormhole solution in the four dimensional Euclidean space does not exist above the compactification scale $\Lambda_{com}$. We should take into account a momentum cut-off at the compactification scale  $1/\Lambda_{com}$ in the estimation of Eq. \eqref{eq:Swh}, which would lead to a large wormhole Euclidean action \cite{Giddings:1987cg}
\begin{equation}\label{eq:Swh-extradimension}
    S_{wh} \simeq \frac{3\pi^2}{8}\frac{M_{\rm PL}^2}{\Lambda_{com}^2}.
\end{equation}
Provided that $F_A\simeq \Lambda_{com}$ we obtain $F_A \leq 2.6\times 10^{17}$ GeV to have a good quality for the quintessence EW axion. Although we can not estimate the shift-symmetry breaking effect above the compactification scale, we consider $F_A \simeq 2.6\times 10^{17}$ GeV is sufficiently small to protect the quintessence axion quality. In this case, the degree of the fine-tuning of the initial condition becomes much milder, i.e., $|{A_I}/{F_A} -\pi| \le 10^{-3}$. Additionally, with $F_A\simeq 2.6\times 10^{17}$ GeV we can explain the cosmic birefringence if the EW axion is rolling now \cite{Choi:2021aze}.

Another viable way is to consider two axions, both of which couple to the strong $SU(3)_C$ and weak $SU(2)_L$ gauge fields~\cite{Kim:2004rp}. We consider the following Lagrangian:
\begin{equation}\label{eq:two-axions}
\begin{split}
\mathcal{L} \supset& \frac{g_2^2}{32\pi^2}
\left(m_1 \frac{A_1}{F_{1}}
+m_2 \frac{A_2}{F_{2}}\right) W_{\mu \nu}^i \widetilde{W}^{i\mu \nu} \\
&~~~~~+\frac{g_3^2}{32\pi^2} \left(n_1 \frac{A_1}{F_{1}}
+n_2 \frac{A_2}{F_{2}}
\right) G_{\mu \nu}^a \widetilde{G}^{a\mu \nu},
\end{split}
\end{equation}
where $m_{1,2}$ and $n_{1,2}$ are anomaly coefficients, and $\bm{G}$ is a gluon field. It is assumed that $F_1 \sim F_2 \sim 10^{16}\,{\rm GeV}$ to allow good qualities for both axions. Assuming $m_1n_2-n_1m_2\neq0$, in the following basis,
\begin{equation}
\left( 
\begin{array}{c}
    A  \\
    A^\prime
\end{array}
\right)
=
\left( 
\begin{array}{cc}
    \cos\beta & -\sin\beta  \\
     \sin\beta & \cos\beta
\end{array}
\right)
\left( 
\begin{array}{c}
    A_1  \\
    A_2
\end{array}
\right)
\end{equation}
where $\tan\beta=\frac{n_1F_2}{n_2F_1}$, only $A'$ couples to $\bm{G} \widetilde{\bm{G}}$ and obtains a large mass through the QCD non-perturbative effects. ({$A'$ is nothing but the QCD axion, solving the strong CP problem.}) Due to the large mass, $A^\prime$ is quickly stabilized to the minimum and can be ignored. The other orthogonal component $A$ remains massless with the following effective potential\footnote{We omit the sign.}
\begin{equation}\label{eq:A-effective}
    \mathcal{L}_{\rm eff} =  \frac{g_2^2}{32\pi^2}\frac{A}{F_{\rm eff}}W_{\mu \nu}^i \widetilde{W}^{i\mu \nu}\,,
\end{equation}
with
\begin{equation}\label{eq:F_eff}
    F_{\rm eff} = \frac{\sqrt{n_1^2F_2^2+n_2^2F_1^2}}{|m_1n_2-m_2n_1|}\,.
\end{equation}
Then, $A$ only couples to $\bm{W}\widetilde{\bm{W}}$ and can be identified as our EW axion.
A larger effective decay constant $F_A = F_{\rm eff} \sim2\times10^{17}$ GeV can be obtained with $n_{1,2}\sim10$ but a small $m_1 n_2 -m_2 n_1$ (e.g., $m_1=m_2=1$, $n_2=10$ and $n_1=9$) for instance.



For possibility (3), the EW axion can starts from a general point between $0$ and $\pi F_A$. Then, the oscillation of the EW axion after $H\sim m_A$ makes 
it become a small fraction of dark matter. In fact, it is encouraging that we can naturally explain the observed cosmic birefringence \cite{Minami:2020odp,Diego-Palazuelos:2022dsq,Eskilt:2022cff} by the axion dark matter oscillation \cite{Lin:2022niw}.\footnote{This is also discussed in \cite{Fujita:2020ecn,Fujita:2020aqt}. However, they assume $F_A = M_{\rm PL}$ and the obtained parameter space is different from that in \cite{Lin:2022niw}. Furthermore, they may have a serious axion quality problem with such a large $F_A$.}

\begin{acknowledgments}
 T.\ T.\ Y.\ deeply thanks Li Fu for her helpful suggestions and encouragement and he is supported in part by the China Grant for Talent Scientific Start-Up Project and by Natural Science Foundation of China (NSFC) under grant No.\ 12175134 as well as by World Premier International Research Center Initiative (WPI Initiative), MEXT, Japan. N. Y. is supported by a start-up grant from Zhejiang University.
\end{acknowledgments}

\providecommand{\href}[2]{#2}\begingroup\raggedright\endgroup

\end{document}